\documentclass[letter,aps,prx,article,twocolumn,floatfix]{revtex4}
\date{\today}
\usepackage{epsfig}
\usepackage{subfigure}
\usepackage{graphicx}
\usepackage{dcolumn}
\usepackage{bm}
\usepackage[colorlinks,citecolor = blue,linkcolor=blue,hyperindex,CJKbookmarks]{hyperref}
\usepackage{float}
\usepackage{hyperref}
\usepackage{comment}
 \usepackage{tensor}
 \usepackage{mathtools}
 \usepackage{calc}
\usepackage{accents}
\newcommand{\eph}{{\it e}-ph}

\begin{document}

\title{Interplay of Kekul\'{e} bond order and lattice instability in $\mathrm{C}_6\mathrm{Li}$}
\author{Yuanhao Zhang}
\affiliation{Department of Physics, Northeastern University, Shenyang 110819, China}
\author{Zi Yuan}
\affiliation{Department of Physics, Northeastern University, Shenyang 110819, China}
\author{Xiangru Kong}
\affiliation{Department of Physics, Northeastern University, Shenyang 110819, China}
\author{Weijiang Gong}
\affiliation{Department of Physics, Northeastern University, Shenyang 110819, China}
\author{Shaozhi Li}
\email{lishaozhi@mail.neu.edu.cn}
\affiliation{Department of Physics, Northeastern University, Shenyang 110819, China}
\date{\today}
\begin{abstract}
Understanding the interplay between charge order and lattice instability in quantum materials remains a central challenge, as their coexistence often obscures causal relationships. This work introduces $\mathrm{C}_6\mathrm{Li}$ as a novel platform to investigate charge order mediated by two distinct mechanisms. We show that the hybridization between carbon $\pi$ and lithium $s$ orbitals generates an effective long-range hopping within Li-centered hexagons. This hopping drives a Kekul\'{e} bond order, whose structure varies with charge density and the sign of the hopping. This bond order induces a Kekul\'{e} lattice distortion via electron-phonon coupling. In the limit where lithium atoms are distant from the graphene layer, a Fermi surface nesting-driven Kekul\'{e} bond order emerges, stabilized by the electron-phonon interaction. Our results establish $\mathrm{C}_6\mathrm{Li}$ as a tunable platform for elucidating the causal hierarchy between electronic and structural orders in quantum materials.
\end{abstract}

\maketitle
{\it Introduction.} The charge density wave (CDW) is a collective quantum phenomenon in which the electron density exhibits spatial modulation, frequently observed in low-dimensional materials. Notable examples include quasi-1D blue bronzes $\text{K}_{0.3}\text{MoO}_3$~\cite{travaglinicharge1984}, two-dimensional (2D) transition metal dichalcogenides~\cite{vescolidynamics1998,nasufermionic2016}, the kagome superconductor $\text{AV}_3\text{Sb}_5$~\cite{niecharge2022,zhongtesting2023}, the cuprate superconductor $\text{La}_{2-x}\text{Ba}_x\text{CuO}_4$~\cite{TranquadaEvidence1995,TranquadaNeutron1996,HuckerStripe2011}, and infinite-layer nickelates  $\text{NdNiO}_2$~\cite{TamCharge2022,KriegerCharge2022}. Unraveling the microscopic origin of the CDW state in these quantum materials remains a central challenge in condensed matter physics.

 \begin{figure*}[t]
\begin{center}
\includegraphics[width=0.9\textwidth]{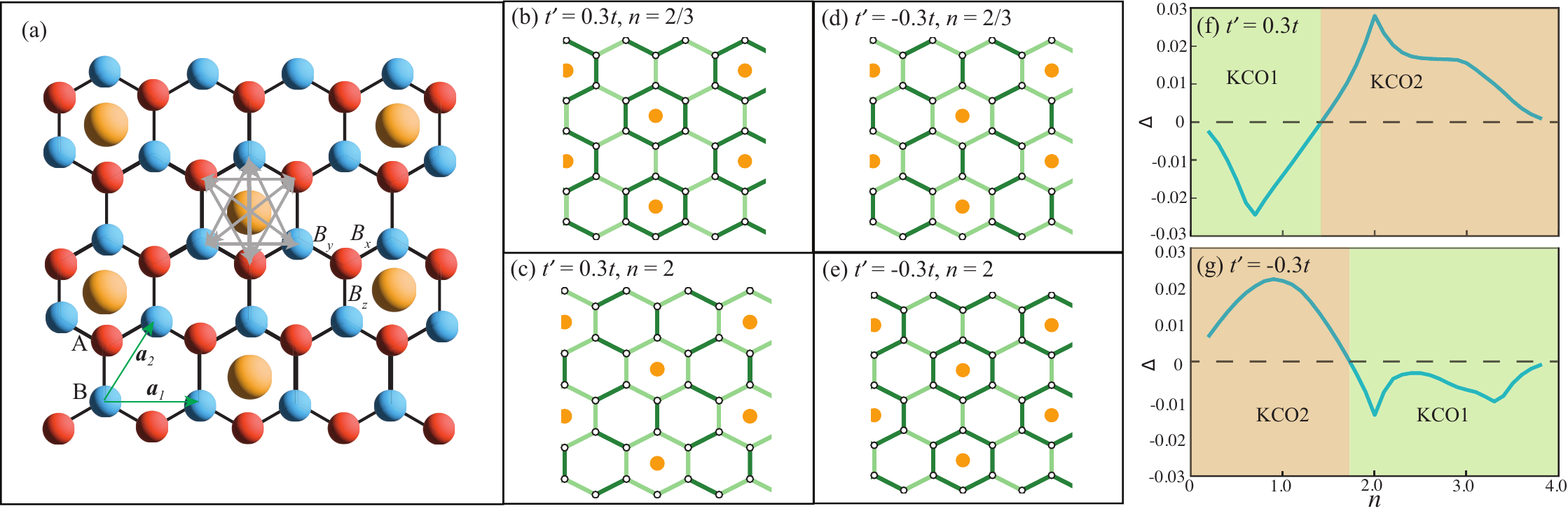}
\caption{ (a) The honeycomb lattice. (b) The nearest-neighbor (NN) charge bond $B_\gamma$ on the honeycomb lattice for $n=2/3$ and $t^\prime=0.3t$. (c) The NN charge bond on the honeycomb lattice for $n=2$ and $t^\prime=0.3t$. (d) The NN charge bond on the honeycomb lattice for $n=2/3$ and $t^\prime=-0.3t$. (e) The NN charge bond on the honeycomb lattice for $n=2$ and $t^\prime=-0.3t$. (f) The Kekul\'{e} bond order $\Delta$ as a function of density $n$ at $t^{\prime}=0.3t$. (g) The Kekul\'{e} bond order $\Delta$ as a function of density $n$ at $t^{\prime}=-0.3t$. 
\label{fig:1} 
}
\end{center}
\end{figure*}

In conventional quantum many-body theory, the CDW state is often associated with Fermi surface nesting facilitated by the electron-phonon (\eph) interaction~\cite{BergerTwo1995, Bauercompetiton2010, NowadnickCompetition2012, WernerPhonon2013, LiCompeting2017, LiSuppressed2023}. According to the Peierls theorem, the {\eph} interaction enhances the charge instability induced by Fermi surface nesting, which in turn leads to lattice distortion and drives the system into a CDW state~\cite{GrunerThe1988}. However, this theorem encounters challenges when applied to many quantum materials, such as the stripe CDW state observed in the cuprate superconductor $\text{La}_{2-x}\text{Ba}_x\text{CuO}_4$ and the kagome superconductor $\text{AV}_3\text{Sb}_5$~\cite{MiaoIncommensurate2018,LinStrongly2020,WangCharge2021,ChristensenTheory2021,YeStructural2022, ChenAbsence2025}. For the cuprate superconductor, numerous numerical studies have demonstrated that the stripe charge order arises from the electron-electron interaction~\cite{BoStripe2017,PeizhiRobust2023}. Consequently, the lattice instability observed in the resonant inelastic x-ray scattering experiment may be a second effect induced by the charge order~\cite{HuangQuantum2021}. In the case of $\text{AV}_3\text{Sb}_5$, multiple mechanisms have been proposed to explain the CDW, including Van Hove singularity scenarios, Fermi surface nesting, sublattice interference, and momentum-dependent {\eph} interaction~\cite{KieselSublattice2012,KieselUnconventional2013,DennerAnalysis2021,KangTwofold2022,LuoElectronic2022,ProfeKagome2024,StefanPhononon2025}. Yet, no consensus has been reached on the dominant mechanism. These examples suggest that a CDW is not necessarily driven by Fermi surface nesting, and the accompanying phonon softening may also originate from the charge order itself or the momentum-dependent {\eph} interaction. However, since CDW formation and phonon softening occur simultaneously, distinguishing their origin is challenging. This complexity fuels ongoing debates about the primary mechanism of CDWs in materials such as $2H\text{-NbSe}_2$, $1T\text{-TiSe}_2$, and $\text{RTe}_3$ \cite{RuEffect2008,WeberExtended2011,HellmannTime2012, MaAmetallic2016, NovkoElectron2022}.

In this work, we propose a platform that can realize a CDW state through two different mechanisms. Specifically, we investigate alkali atoms on a graphene compound, i.e., $\text{C}_6\text{Li}$, where alkali atoms form a $(\sqrt{3}\times\sqrt{3})R30^{\circ}$ triangular lattice on the two-dimensional honeycomb lattice, as illustrated in Fig.~\ref{fig:1}(a)~\cite{CalandraTheoretical2005,CalandraElectronic2007,farjamenergy2009,HazratiLi2014,WangVan2014,YangDensity2016,BaoExperimental2021,Liessential2022,HollandAb2022}. Angle-resolved photoemission spectroscopy and scanning tunneling microscopy studies show that depositing lithium atoms on top of the monolayer graphene breaks chiral symmetry, inducing a gapped Kekul\'{e} bond order~\cite{BaoExperimental2021}. Early theoretical work proposed that a tight-binding model with anisotropic nearest-neighbor hoppings on the honeycomb lattice can generate such a gapped Kekul\'{e} state. However, that study did not identify the microscopic origin of the anisotropy or the underlying mechanism driving the Kekul\'{e} bond order in $\text{C}_6\text{Li}$~\cite{farjamenergy2009}.

In contrast, our study proposes a different theoretical model that explains both the emergence of the gapped Kekul\'{e} state and the role of phonons. We show that the Kekul\'{e} bond order originates from an effective long-range hopping that is induced by hybridization between the $\pi$ orbital of carbon atoms and the $s$ orbital of alkali atoms. At charge density $n=2$, this long-range hopping drives the system into an insulating Kekul\'{e} state. At $n=2/3$, we identify a phase transition from a metallic to an insulating Kekul\'{e} state as the long-range hopping strength increases. 
In addition, we observe that the metallic Kekul\'{e} state persists across all doping levels, with the exception of $n=2$ and $n=2/3$. Through analytical analysis, we demonstrate that this long-range hopping term induces a scattering between electronic states with momenta $\mathbf{k}$ and $\mathbf{k} + \mathbf{Q}_K$, where $\mathbf{Q}_K$ is the wave vector associated with the bond order. Furthermore, we examine the effect of phonons and observe that the Kekul\'{e} bond order induces a corresponding lattice distortion, which in turn enhances the Kekul\'{e} bond order. These results show that the CDW state, induced by the electronic scattering, drives the lattice instability in $\text{C}_6\text{Li}$. When the lithium atom is far from the graphene layer, the effective long-range hopping term becomes negligible. In this limit, a sufficiently strong {\eph} interaction induces the Kekul\'{e} bond order at $n=2/3$ and $n=2$. We attribute this bond order to the Fermi surface nesting assisted by the {\eph} coupling. These findings suggest that our system serves as a new platform for exploring the interplay between charge order and lattice instability.

{\it Model.} To simulate the low-energy physics of $\text{C}_6\text{Li}$, we consider a single-band tight-binding model on the honeycomb lattice. The effect of the alkali atom is modeled by introducing a long-range hopping term between $\pi$ electrons on carbon atoms. This long-range hopping arises from the hybridization between carbon $\pi$ electrons and the $s$ orbital electrons of alkali atoms~\cite{Guzmansuperconductivity2014}. Due to the isotropy of the $s$ orbital, we treat the second and third nearest-neighbor hoppings within each hexagon containing a centered alkali atom as equal, as shown in Fig.~\ref{fig:1}(a). The Hamiltonian of this tight-binding model is then given by 
\begin{align}
H_0=-t\sum_{\langle i,j \rangle, \sigma} c_{i,\sigma}^{\dagger}c_{j,\sigma}^{\phantom\dagger} - t^\prime \sum_{i,j\in P} c_{i,\sigma}^{\dagger}c_{j,\sigma}^{\phantom\dagger} -\mu\sum_{i,\sigma}n_{i,\sigma}.
\end{align}
Here, $c_{i,\sigma}$ ($c_{i,\sigma}^{\dagger}$) is the annihilation (creation) operator for electrons at site $i$ with spin $\sigma$. $t$ describes the nearest-neighbor hopping, and $\mu$ is the chemical potential, which controls the charge density. $t^\prime$ describes the long-range hopping within a hexagon containing a centered alkali atom, and $P$ denotes the set of all such hexagons. We note that our model differs from the model proposed in Ref.~\cite{farjamenergy2009}, in which two different nearest-neighbor hoppings were adopted to generate the Kekul\'{e} bond order and the band gap. Throughout this work, we use $t$ as the unit of energy.

\begin{figure*}[t]
\begin{center}
\includegraphics[width=0.99\textwidth]{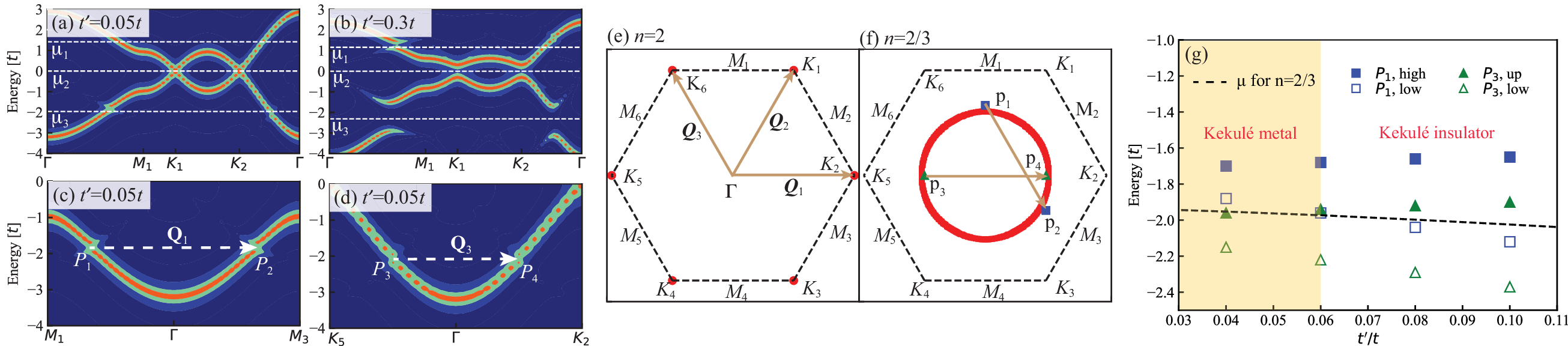}
\caption{\label{fig:2} (a) The spectral function along the high-symmetry path for $t^\prime=0.05t$. (b) The spectral function along the high-symmetry path for $t^\prime=0.3t$. (c) The spectral function for $t^\prime=0.05t$ along the $M_1$-$\Gamma$-$M_3$ path. (d) The spectral function for $t^\prime=0.05t$ along the $K_5$-$\Gamma$-$K_2$ path. 
(e) The Fermi surface at $n=2$ and $t^\prime=0.05t$. (f) The Fermi surface at $n=2/3$ and $t^\prime=0.05t$. (g) The phase transition between the metallic and insulating Kekul\'{e} bond orders as a function of $t^\prime$ at $n=2/3$.
}
\end{center}
\end{figure*}

{\it O-type Kekul\'{e} bond order.} The long-range hopping inside the hexagon containing a centered Li atom naturally induces the O-type Kekul\'{e} bond order. Figures~\ref{fig:1}(b) and~\ref{fig:1}(c) display the nearest-neighbor (NN) charge bond configurations at fillings $n=2/3$ and $n=2$, respectively, for $t^\prime=0.3t$.
The NN charge bond is defined as $B_{\gamma}=\langle c_{{\bf r},\mathrm{A}}^{\dagger}c_{{\bf r}+{\bf r}_\gamma,\mathrm{B}}\rangle$, where $\gamma$ denotes $x$, $y$, and $z$ as shown in Fig.~\ref{fig:1}(a). Here, ${\bf r}_x={\bf a}_2$, ${\bf r}_y=-{\bf a}_1+{\bf a}_2$, and ${\bf r}_z=0$, where ${\bf a}_1$ and ${\bf a}_2$ are the lattice vectors, each with length $a$ as shown in Fig.~\ref{fig:1}(a). 
At $n=2/3$ and $t^{\prime}=0.3t$, the NN charge bond on the hexagon with a centered Li atom is stronger than those in other hexagons. In contrast, at $n=2$ and $t^{\prime}=0.3t$, the bond associated with the Li-centered hexagon becomes weaker than the others. These two distinct bond order patterns correspond to different real-space charge distributions, which we label as KCO1 (for $n=2/3$) and KCO2 (for $n=2$). Interestingly, when the sign of long-range hopping is reversed to $t^\prime=-0.3t$, the bond orders are reversed: KCO2 emerges at $n=2/3$, whereas KCO1 appears at $n=2$.

To understand how the bond order evolves with electron density, we plot the Kekul\'{e} bond order parameter $\Delta$ as a function of the density $n$ for $t^\prime=0.3t$ and $t^\prime=-0.3t$ in Fig.~\ref{fig:1}(f) and Fig.~\ref{fig:1}(g), respectively. The order parameter $\Delta$ is defined as $\Delta=\frac{1}{N}\sum_{r}\mathrm{e}^{i{\bf Q}_K\cdot {\bf r}}B_y({\bf r})$, where ${\bf Q}_K$ is one of the wave vectors $\mathbf{Q}_1$, $\mathbf{Q}_2$, or $\mathbf{Q}_3$, given by: $\mathbf{Q}_1 = \frac{2\pi}{a}(\frac{2}{3\sqrt{3}}, 0)$,
$\mathbf{Q}_2 = \frac{2\pi}{a}(\frac{1}{3\sqrt{3}},\frac{1}{3})$,
$\mathbf{Q}_3 = \frac{2\pi}{a}(-\frac{1}{3\sqrt{3}},\frac{1}{3})$. The Kekul\'{e} bond order is identical for these three wave vectors. Our results reveal a phase transition from KCO1 to KCO2 at $n=1.5$ for $t^\prime=0.3t$, whereas for $t^\prime=-0.3t$, the transition from KCO2 to KCO1 occurs around $n=1.8$. More results for other values of $t^\prime$ are shown in the Supplemental Material~\cite{SM}. The experiment in Ref.~\cite{BaoExperimental2021} observed a KCO1 in $\text{C}_6\text{Li}$ near $n=2$, implying that the long-range hopping is negative in $\text{C}_6\text{Li}$.

{\it Coexistence of the Kekul\'{e} bond order and the metallic state.} Although the Kekul\'{e} bond order is robust, band gaps only appear at specific densities. To investigate the formation of the band gap, we study the spectral function along a high-symmetry path for $t^\prime=0.05t$ and $t^\prime=0.3t$, respectively. The high-symmetry path used here is labeled in Fig.~\ref{fig:2}(e). The white dashed lines in Figs.~\ref{fig:2}(a) and~(b) from top to bottom represent the chemical potentials for $n=10/3$, $n=2$, and $n=2/3$, labeled as $\mu_1$, $\mu_2$, and $\mu_3$, respectively. The long-range hopping breaks particle-hole symmetry, resulting in chemical potentials for $n=10/3$ and $n=2/3$ that are not symmetric about zero energy. At $t^\prime=0.3t$, two band gaps form at $\mu_2$ and $\mu_3$, indicating an insulating state at $n=2/3$ and $n=2$. Near $\mu_1$, there is one band gap along the $\Gamma$-$M_1$ path; however, the lower band edge along the $K_2$-$\Gamma$ path is higher than $\mu_1$, suggesting a pseudogap at $n=10/3$. This pseudogap also exists at $n=2/3$ when $t^\prime$ is small (i.e., $t^\prime=0.05t$). We observe that the band gap does not exist at other densities, indicating that the Kekul\'{e} bond order can coexist with a metallic state.

{\it Electronic-scattering–driven charge order.} To elucidate the origin of the band gap and the charge order induced by the long-range hopping in the Li-centered hexagon, we perform an analytical analysis (see details in the Supplemental Material~\cite{SM}). Our analysis shows that the long-range hopping term in the Li-centered hexagon can be transformed into a momentum-space coupling between electronic states at $\mathbf{k}$ and $\mathbf{k}+\mathbf{Q}_K$, where $\mathbf{Q}_K$ is the wave vector associated with the bond order. This coupling corresponds to scattering between electronic states of the same energy that are separated by $\mathbf{Q}_K$ in the momentum space, which directly breaks translational symmetry by $\mathbf{Q}_K$ and induces the bond order and a band gap. At filling $n=2$, the Fermi surface of graphene consists of six Dirac points, and two nearest-neighbor Dirac points are separated by one of the wave vectors $\mathbf{Q}_1$, $\mathbf{Q}_2$, or $\mathbf{Q}_3$, as illustrated in Fig.~\ref{fig:2}(e). Consequently, scattering between two nearest-neighbor Dirac points leads to the opening of a band gap at the Fermi surface, which increases linearly with $t^\prime$. At filling $n=2/3$, the Fermi surface exhibits a single hole pocket centered at the $\Gamma$ point, as shown in Fig.~\ref{fig:2}(f). The band gap along the $\Gamma$-$M_1$ path shown in Fig.~\ref{fig:2}(c) arises from the scattering between electrons at points $P_1$ (located at $\frac{2}{3}\Gamma M_1$) and $P_2$ (located at $\frac{2}{3}\Gamma M_2$), as shown in Fig.~\ref{fig:2}(f). Similarly, the band gap along the $\Gamma$-$K_2$ path shown in Fig.~\ref{fig:2}(d) originates from the scattering between electrons at $P_3$ (located at $\frac{1}{2}\Gamma K_2$) and $P_4$ (located at $\frac{1}{2}\Gamma K_5$) points, depicted in Fig.~\ref{fig:2}(f). Notably, $P_1$ and $P_2$ lie outside the hole pocket, whereas $P_3$ and $P_4$ lie inside. Consequently, the eigenenergy at $P_1$ is slightly higher than that at $P_3$, as is evident from comparing Figs.~\ref{fig:2}(c) and ~\ref{fig:2}(d).
Figure~\ref{fig:2}(g) displays the eigenenergies (below zero energy) at points $P_1$ and $P_3$ as a function of $t^\prime$.
The black dashed line denotes the chemical potential for $n=2/3$. As $t^\prime$ increases, the lower branch of the eigenmodes at $P_1$ shifts downward and crosses the Fermi level at approximately $t^\prime=0.06t$. Meanwhile, the upper branch of the eigenmodes at $P_3$ shifts upward and crosses the Fermi level at a comparable value of $t^\prime$. These results indicate that a phase transition from the metallic Kekul\'{e} state to the insulating Kekul\'{e} state occurs at $t^\prime=0.06t$. 

\begin{figure*}[t]
\begin{center}
\includegraphics[width=0.7\textwidth]{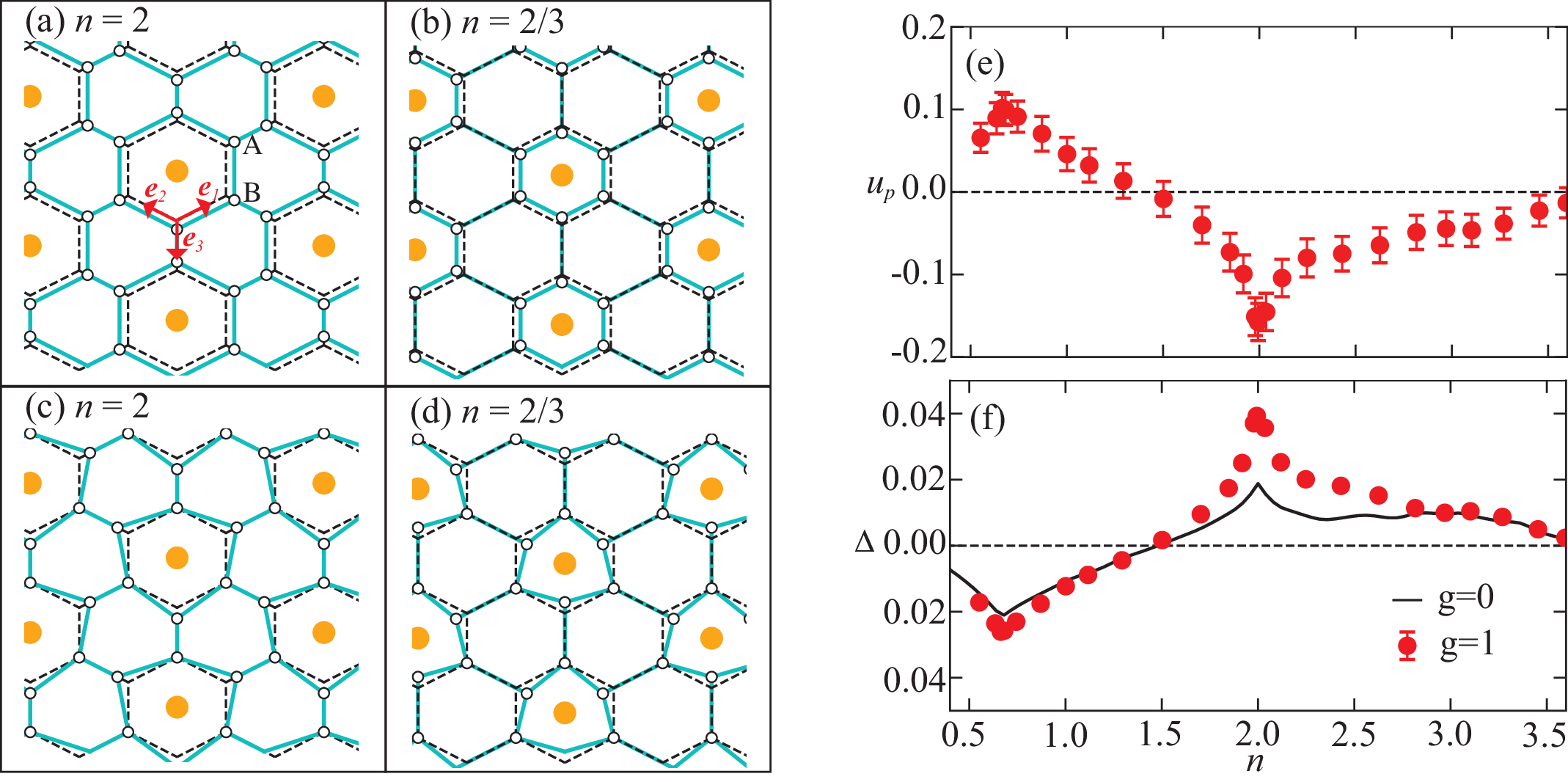}
\caption{\label{fig:3} Lattice distortion at $n=2$ and $n=2/3$ for $t^\prime=0.2t$. (a),(b) Lattice distortion patterns in the honeycomb lattice, where both A and B sublattices are allowed to move. (c),(d) Lattice distortion patterns in the honeycomb lattice, where only the A sublattice is allowed to move. (e) The projected displacement as a function of density $n$ at $g=1$, $t^\prime=0.2t$, and $\beta=10t$. (f) The Kekul\'{e} bond order as a function of density at $t^\prime=0.2t$ and $\beta=10t$. The phonon frequency is set as $\Omega=t$.
}
\end{center}
\end{figure*}

{\it Electron-phonon interaction.} Next, we study the impact of phonons on this bond order. Here, we consider the phonon Hamiltonian with the NN interaction, which is defined as 
\begin{eqnarray}
H_\mathrm{ph}&=&M\sum_{i}\left(\frac{\dot{{\bf u}}_{i,A}^2}{2}+\frac{\dot{{\bf u}}_{i,B}^2}{2}\right)+\frac{K}{2}\sum_{\langle i, j \rangle} |{\bf r}_{i,A}+{\bf u}_{i,A} \nonumber\\
&&-{\bf r}_{j,B}-{\bf u}_{j,B}|^2, 
\end{eqnarray}
where $\langle i,j \rangle$ represents the NN carbon pairs. $M$ is the carbon mass, and $K$ is the spring constant. ${\bf r}_{i,\alpha}$ ($\alpha=$A, B) denotes the equilibrium position of a carbon atom on sublattice $\alpha$ in unit cell $i$, and $\mathbf{u}_{i,\alpha}$ is its displacement from equilibrium. When carbon atoms move away from their equilibrium position, the distance between carbon atoms is changed, which modifies the electron hopping between two sites. To describe this effect, we set the {\eph} coupling as 
\begin{align}
H_\mathrm{eph} = -g t \sum_{\langle i,j \rangle,\sigma} \left( \lvert \mathbf{r}_{i,A}+\mathbf{u}_{i,A} - \mathbf{r}_{j,B}-\mathbf{u}_{j,B} \rvert - a \right) c_{i,\sigma}^{\dagger}c_{j,\sigma}^{\phantom\dagger},
\end{align}
where $a$ is the equilibrium bond length, set to 1, and $g$ denotes the {\eph} interaction strength.

To understand the vibration mode of each carbon atom, we perform a mean-field analysis of a $2\times 48\times 48$ lattice for $t^\prime=0.2t$ and $K=t$. Figures~\ref{fig:3}(a) and~\ref{fig:3}(b) plot the displacement of each atom at $n=2$ and $n=2/3$, respectively.  The black dashed line denotes the original honeycomb lattice, and the black circles represent carbon atoms, which are connected via a cyan line. Interestingly, the vibration mode at $n=2$ differs from that at $n=2/3$. At $n=2$, the hexagon with a centered Li atom expands, while this hexagon shrinks at $n=2/3$. This difference in vibration modes is attributed to different Kekul\'{e} bond orders at $n=2$ and $n=2/3$, as shown in Fig.~\ref{fig:1}(b) and Fig.~\ref{fig:1}(c). 
The vibration mode of the A atoms when the B sublattice is fixed is consistent with the case where both sublattices are free to vibrate (Figs.~\ref{fig:3}(c) and~\ref{fig:3}(d)). These results imply that the lattice distortion is driven by the charge order. The same conclusion can also be obtained from results for $t^\prime=-0.2t$ (see details in the Supplemental Material~\cite{SM}).

 To further explore the relationship between the charge and lattice orders, we perform simulations using the Determinant Quantum Monte Carlo on a $2\times 6\times 6$ lattice with a phonon frequency $\Omega=\sqrt{K/M}=t$~\cite{LiQuantum2020, CaiRobustness2022, LiSuppressed2023,CaiHigh2025}. To reduce the computational cost, we only allow the A sublattice to vibrate in the plane. Figure~\ref{fig:3}(e) plots the projected displacement $u_p$ as a function of the density $n$ at $g=1$, $t^\prime=0.2t$ and $\beta=10/t$. The projected displacement is defined as
 \begin{align}
u_p = \frac{1}{N}\sum_{i} {\bf u}_i \cdot {\bf e}_i,
 \end{align}
 where ${\bf e}_i$ is a unit vector along the direction of $B_x$, $B_y$, and $B_z$ bonds, which is determined by the direction of motion of each A atom shown in Fig.~\ref{fig:3}(d). We find that the projected displacement $u_p$ is positive when $n<0.75$, and it becomes negative when $n>0.75$, implying that a phase transition between KCO1 and KCO2 occurs at $n=0.75$. This phase transition can also be observed from the behavior of the Kekul\'{e} order $\Delta$, shown in  Fig.~\ref{fig:3}(f). Compared to the Kekul\'{e} order at $g=0$, $\Delta$ at $g=1$ is larger, implying that the {\eph} interaction enhances the bond order, even though this interaction is not its primary cause.

\begin{figure}[t]
\begin{center}
\includegraphics[width=0.99\columnwidth]{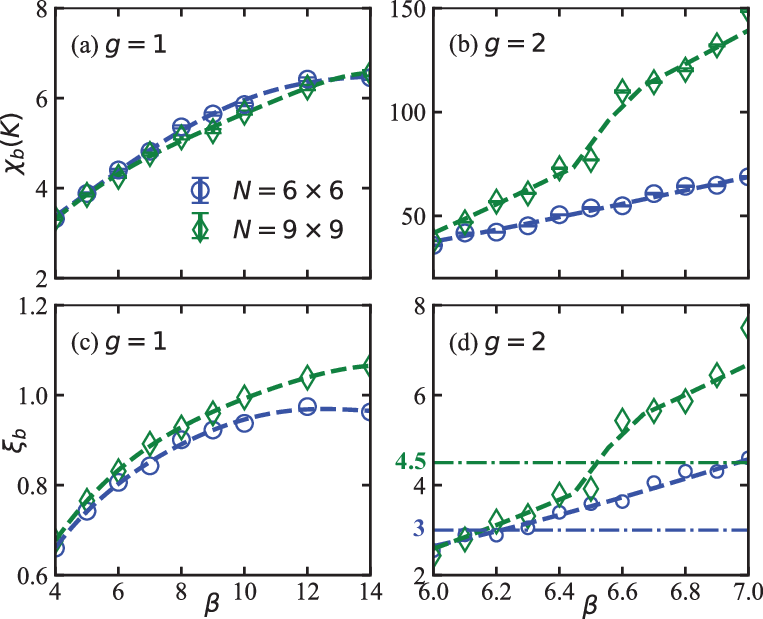}
\caption{\label{fig:4} Bond correlation functions $\chi_b(K)$ at the K momentum point for (a) $g=1$ and (b) $g=2$.  The correlation length $\xi_b$ of $\chi_b(k)$ for (c) $g=1$ and (d) $g=2$. The phonon frequency is set as $\Omega=t$, and the charge density $n$ is 2/3. The dashed curves are guides to the eye.}
\end{center}
\end{figure}

{\it Fermi surface nesting driven charge order.} Our study reveals that the long-range hopping in the $\mathrm{C}_6\mathrm{Li}$ structure induces charge order, which in turn drives lattice instability. This raises the question of whether the {\eph} interaction can induce Kekul\'{e} bond order when lithium atoms move far away from the graphene layer. 
It has been established that the Su-Schrieffer-Heeger (SSH) {\eph} interaction can generate a Kekul\'{e} bond order in graphene at $n=2$ \cite{MalkarugeKekule2024}. Here, we extend this investigation to the $n=2/3$ filling regime.

Figure~\ref{fig:4} presents the bond correlation functions $\chi_b(K)$ at the K point and the corresponding correlation lengths $\xi_b$ for coupling strengths $g=1$ and $g=2$. (See Supplemental Material for definitions of $\chi_b(K)$ and $\xi_b$~\cite{SM}.) For $g=1$, $\chi_b(K)$ exhibits negligible system-size dependence, and both $\chi_b(K)$ and $\xi_b$ increase as the temperature decreases. Although the values of $\chi_b(K)$ for $N=6\times6$ and $N=9\times9$ are nearly identical, the correlation length $\xi_b$ is slightly larger for the $N=9\times9$ system. Importantly, $\xi_b$ remains below half of the lattice size at low temperatures for both system sizes, indicating the absence of long-range bond order. In contrast, at $g=2$, $\chi_b(K)$ for $N=9\times 9$ increases more rapidly with decreasing temperature than for $N=6\times 6$. The corresponding correlation lengths exceed half the lattice length at $\beta=6.2$ ($N=6\times 6$) and $\beta=6.55$ ($N=9\times 9$), confirming the emergence of bond order at low temperatures. The shift of the transition temperature with system size is attributed to the finite-size effect. Furthermore, by analyzing the noninteracting susceptibility of graphene (see details in the Supplemental Material~\cite{SM}), we find peak structures at the K point for $n=2$ and $n=2/3$, indicating that the Kekul\'{e} bond order originates from Fermi surface nesting, assisted by the {\eph} interaction. 

{\it Conclusion.}  Our study not only explains the microscopic mechanism of the bond order in $\mathrm{C}_6\mathrm{Li}$ but also provides a platform to elucidate the interplay between charge order and lattice instability. First, when lithium atoms are close to the graphene layer, hybridization between the $2p_z$ orbital of carbon atoms and the $s$ orbital of lithium atoms generates an effective long-range hopping between electrons on the $\pi$ orbitals in the Li-centered hexagon, which induces a Kekul\'{e} bond order. The structure of the Kekul\'{e} bond order depends on the charge density and the sign of the long-range hopping. We demonstrate that this electronically driven bond order subsequently induces a lattice instability. Second, when lithium atoms are far from the graphene layer, a strong in-plane SSH e-ph interaction can stabilize the Kekul\'{e} lattice distortion by coupling to electronic states connected by Fermi surface nesting. In $\text{C}_6\text{Li}$, the intrinsic SSH e-ph interaction may be too weak to stabilize this distortion. However, external in-plane pressure can be applied to enhance the e-ph interaction, thereby promoting this nesting-driven mechanism. Therefore, our work demonstrates that $\text{C}_6\text{Li}$ is a tunable platform that can realize a CDW state through two different mechanisms.

{\it Acknowledgement.} We thank Xianxin Wu and Jun Zhan for insightful discussions. This work was supported by the National Natural Science Foundation of the People's Republic of China (Young Scientists Fund, Grant No. 12204236). The authors also acknowledge the startup funding support from Northeastern University, Shenyang.

{\it Data Availability.} The data that support the findings of this article are not publicly available upon publication because it is not technically feasible and/or the cost of preparing, depositing, and hosting the data would be prohibitive within the terms of this research project. The data are available from the authors upon reasonable request.

\bibliography{main}

\end{document}